\documentstyle[12pt,a4]{article}
\newcommand{\be}{\begin{equation}}
\newcommand{\ee}{\end{equation}}
\begin{document}
\vspace{20.0mm}
\title{\bf
\Large
Bound states in a nonlinear Kronig Penney model\\}

\author{{\Large Stavros Theodorakis and Epameinondas Leontidis}\\
         \\
            Dept. of Natural Sciences, University of Cyprus\\  
            {\it P.O. Box 537, Nicosia 1678, Cyprus}}

\maketitle

\vspace{35.0mm}
\begin{abstract}

\noindent 
We study the bound states of a Kronig Penney potential
for a nonlinear one-dimensional Schrodinger equation. This
potential consists of a large, but not necessarily infinite,
number of equidistant $\delta$-function wells. We show
that the ground state can be highly degenerate. Under certain
conditions furthermore, even the bound state that would be
normally the highest can have almost the same energy as the
ground state. This holds for other simple periodic potentials as
well. 

\end{abstract}

\newpage
\section{Introduction}
 
In this paper we shall study an unusual generalization of the
one-dimensional Kronig Penney model. We shall examine in
particular the spectrum of the bound states for a Kronig Penney
potential V(z), having added though a nonlinear term
to the Schrodinger equation. Our arguments will be valid in the
case of other simple periodic potentials as well. 

\vspace*{3mm}
Such nonlinear equations with periodic potentials arise in the
Ginzburg-Landau treatment of various phenomena in condensed
matter physics. In layered superconductors for example, such as
the high temperature ones, a periodic potential such as the
Kronig Penney potential can describe the periodically modulated
superconductivity of the samples [1]. Spatially varying
parameters in the nonlinear Schrodinger equation were also used
to describe the periodic variation of the impurity concentration
in superconductors [2], high T$_{c}$ Josephson field effect
transistors [3], as well as grain boundaries in superconducting
bicrystals [4], while nonlinear Kronig Penney models were
used for studying twinning-plane superconductivity [5]. The
nonlinear Schrodinger equation must be used in order to describe
all these various phenomena, including the relevant phase
transitions. The nonlinear Schrodinger equation has been studied
repeatedly, but mostly in regard to its solitons [6], and usually
for nonperiodic potentials. In this work the emphasis is placed
on studying the bound states, rather than solitons. 

\vspace*{3mm}
We shall study the excited states for the equation
\be
-\frac{\hbar^{2}}{2M}\frac{\partial^{2}\Psi}{\partial z^{2}}
+V(z)\Psi+\beta|\Psi|^{2}\Psi = 0.
\ee
The nonlinear term forbids the arbitrary normalization
of $\Psi$.

\vspace*{3mm}
The potential we have in mind is a Kronig-Penney potential, but
it could be in general any simple oscillatory potential. In this
work we choose
\be
V(z)=V_{0}\Bigl[1-\alpha\sum_{n}\delta(\frac{z}{d}-n-
\frac{1}{2})\Bigr],
\ee
with $\alpha$ and V$_{0}$ positive. The crucial parameter in this
potential is the periodicity length d. The number of
wells is large, but not necessarily infinite.

\vspace*{3mm}
Equation (1.1) minimizes the energy functional
\be
\int\,\,dz\Bigl[V(z)|\Psi|^{2}
+\beta|\Psi|^{4}/2
+\frac{\hbar^{2}}{2M}\bigg|\frac{\partial\Psi}{\partial
z}\bigg|^{2}\Bigr] . 
\ee
For $M\rightarrow\infty$ we would have $|\Psi|^{2}=-V(z)/\beta$,
in which case $|\Psi|^{2}$ would follow the periodicity of V(z).
If the nonlinear term is omitted, the usual Kronig Penney model
is recovered. In that limit $-V_{0}$ is the energy, and $\alpha
V_{0}$ is the strength of each attractive delta function.

\vspace*{3mm}
We can write the energy functional in dimensionless form,
by measuring z in units of d, the distance between successive
spikes of the potential, $\Psi$ in units of $\sqrt{V_{0}/\beta}$,
and the energy in units of d$V_{0}^{2}/\beta$, where $V_{0}$
is the positive constant that appears in Eq. (1.2), and has the
dimensions of $V(z)$. This constant is taken out of
$V(z)$, so as to render it dimensionless. In other words, 
$V(z)/V_{0}=$ u(z), where u(z) is dimensionless. If we define
then the dimensionless parameter $\nu=\hbar^{2}/2MV_{0}$d$^{2}$,
the energy functional takes the dimensionless form
\be
\int\,dz
\Bigl[u(z)|\Psi|^{2}
+|\Psi|^{4}/2+\nu|\frac{\partial\Psi}{\partial z}|^{2}\Bigr]. 
\ee

\vspace*{3mm}
Note that when the quartic term is omitted, we recover the usual
linear Kronig Penney model, with energy E=$-\hbar^{2}/2M\nu
d^{2}$. In this case the energy values can be found only after
imposing periodic boundary conditions on $|\Psi|^{2}$. There are
then only
certain allowed values of $\nu$, for
a given value of $\alpha$. The size of the wavefunction is
determined by the normalization, and when we minimize the energy
functional under this constraint, we find the energy
eigenvalues, i.e. the minima of the energy functional. 

\vspace*{3mm}
In the nonlinear case on the other hand, the size of $\Psi$ is
determined by the nonlinear terms, through the unconstrained
minimization of the functional of Eq. (1.4). These nonlinear
terms determine fully the behaviour of $\Psi$, without any need
for boundary conditions. In fact, a periodic u(z) will give a
periodic $|\Psi|^{2}$. Furthermore, the parameters $\alpha$ and
$\nu$
are now independent, and for any pair of values of $\alpha$ and
$\nu$ we can find a solution $\Psi$, as long as $\alpha$ is
sufficiently large. We shall see later what is precisely the
lower bound on $\alpha$. The energy of each state will be simply
the value of the energy functional (1.4) at its minimum.
 
\vspace*{3mm}
We see from Eq. (1.4) that for $\nu\rightarrow 0$, when the
potential is very strong, or very weakly periodic, we get
$|\Psi|^{2}\rightarrow$$-$u(z).
Thus $\Psi$ follows very closely the periodicity of the
structure, since it can change very abruptly. In this limit
though the sign of $\Psi$ is arbitrary. So if the spikes of the
potential are very far apart (d is long),
the sign of $\Psi$
could be positive or negative at each spike (see
Fig. 1a).

\vspace*{3mm}
Let us now switch on slowly the parameter $\nu$, bringing the
teeth of the potential comb closer together. Then the
wavefunction between
neighboring spikes could have two forms. If the wavefunction on
two successive spikes A and B is positive, say, then the
wavefunction
in the intervening region will be reduced, and it will go through
a positive minimum value, remaining always positive though (Fig.
1b). If however the wavefunction changes sign in going from
spike B to spike C, then it must pass through a point halfway
between the spikes where it is exactly zero (see Fig. 1b). Since
the wavefunction $\Psi(z)$ minimizes the functional of Eq.
(1.4), the energy equals 
$-\int\,dz\,|\Psi|^{4}/2$, as can be deduced by combining the
dimensionless forms of Equations (1.1) and (1.3). Consequently
the wavefunction has less
energy if it does not go through zero, maintaining always the
same sign. Indeed, in that case the minimum
of $|\Psi|^{4}/2$ is not zero, and hence the area under
$|\Psi|^{4}/2$ is greater.

\vspace*{3mm}
It seems therefore more favorable for the wavefunction to have
the same sign on all spikes of the potential. We say that the
ground state is a
$uniformly$ $positive$ state then. If the spikes of the potential
are too far from each other however,
then the minimum value of the wavefunction between them is
practically zero, and in that case the $uniformly$ $positive$
state
(where
$\Psi$ has the same sign at all spikes) becomes degenerate in
energy
with states that may have $\Psi$ take on negative values at some
spikes, and positive values at others.

\vspace*{3mm}
We can, for example, have a state that is infinitesimally higher
in energy compared to the uniformly positive ground state, and
hence
practically equally preferable, even for spikes not too far
apart.
This state, with $\Psi(0)=0$, $\Psi(z)>0$ when $z>0$, and
$\Psi(z)<0$ when $z<0$, connects regions of different signs of
the wavefunction (see Fig. 2). Then in the intermediate region
$\Psi$ has to go
through zero, and we get a region that reminds us of a domain
wall. For a potential with $infinitely$ many spikes, the
energetically
costly root of $\Psi$ occurs only once, and hence the energy of
this state is equal to the energy of the uniformly positive
ground state.   

\vspace*{3mm}
There can be bound states of Equation (1.1) therefore that are
degenerate to the
ground state, not being everywhere positive. It is the
purpose of this paper to study such bound states, first through
a general variational model (Section 2), and then through an
exact study of the Kronig-Penney potential (Section 3), as well
as through a numerical study of a periodic potential with
gaussian wells (Section 4). We summarize our conclusions in
Section 5.

\section{Variational Study}

In this section we shall examine the possibility of having $\Psi$
change sign in going from one spike of the potential to the next,
as well as the
possibility of having $\Psi$ keep the same sign on neighboring
spikes. In the first case $\Psi$ is odd with respect to the
midpoint between the two spikes, while in the second case it is
even. An arbitrary state of the system will then be a combination
of even and odd pieces. In other words, $\Psi$ will be even
between certain neighboring spikes of the potential, and odd
between others. Thus
$\Psi$ will
maintain its sign between some spikes, and it will change sign
between others. For example, in Fig. 1b $\Psi$ is odd in one
interval, and even in the other two, while in Fig. 2 it is even
everywhere, except for the interval at the center. 

\vspace*{3mm}
We have assumed that the spikes of the potential are at
$z=n+\frac{1}{2}$, where
$n$ is any integer. Let us examine then the two neighboring
quantum wells at the ends of the interval [$n-\frac{1}{2}$,
$n+\frac{1}{2}$]. We adopt
the following odd and even trial wavefunctions, with
respect
to the midpoint ($z=n$), defined on the interval 
[$n-\frac{1}{2}$,$n+\frac{1}{2}]$:
\be
\Psi_{on}(z)=\pm\psi\frac{\sinh[\gamma(z-n)]}{\sinh(\gamma/2)},
\ee
\be
\Psi_{en}(z)=\pm\psi\Biggl[
\frac{\cosh[\gamma(z-n)]}{\cosh(\gamma/2)}-
{\rm sech}^{2}(\gamma/2)\Biggr]\coth^{2}(\gamma/2),
\ee
where $\psi$ and $\gamma$ are variational parameters. We note
that $\Psi_{on}(n)=0$,
$\Psi_{on}(n+{1\over 2})$
$=-\Psi_{on}(n-{1\over 2})=\pm\psi$, and $\Psi_{on}^{\prime}$
$(n+{1\over 2})=\Psi_{on}^{\prime}(n-{1\over 2})$
$=\pm\psi\gamma\coth(\gamma/2)$. Similarly $\Psi_{en}(n)=$
$\pm\psi[\cosh(\gamma/2)-1]/\sinh^{2}(\gamma/2)$,
$\Psi_{en}(n+{1\over 2})=$$\Psi_{en}(n-{1\over 2})=\pm\psi$, and
$\Psi^{\prime}_{en}(n+{1\over 2})=$$-\Psi^{\prime}_{en}$
$(n-{1\over 2})=\pm\psi\gamma\coth(\gamma/2)$.

\vspace*{3mm}
These wavefunctions are such that they can be joined together
in any order to form a continuous wavefunction everywhere,
consisting of even and odd pieces. We could have, for example,
$\Psi=\Psi_{on}$ in [$n-{1\over 2},n+{1\over 2}]$,
$\Psi=\Psi_{e,n+1}$ in [$n+{1\over 2},n+{3\over 2}$],
$\Psi=-\Psi_{o,n+2}$ in [$n+{3\over 2},n+{5\over 2}$], etc.
Furthermore, regardless of the order in which the even and odd
pieces are connected, the slope of the wavefunction is
symmetric around the spikes of the potential.

\vspace*{3mm}
The state with the lowest energy would consist of a chain of even
pieces, because, unlike the odd pieces which have a root at the
midpoint, the even pieces are nowhere equal to zero. Thus the odd
pieces have a higher $-\int\,dz\,|\Psi|^{4}/2$,
which is the exact energy if $\Psi$ is an exact solution of
the equations that minimize the energy functional.

\vspace*{3mm}
The next lowest energy would correspond to the state with only
one odd
piece. This is the state of Fig. 2. It is presumed here though
that the change from the chain of negative pieces to the chain
of positive pieces occurs within just one spacing. The
circumstances under which this will happen will be explored
later. 

\vspace*{3mm}
The state mentioned above is followed by the
state with two odd pieces, and so on, up to the highest state,
which has only odd pieces. In fact, if F$_{e}$ and F$_{o}$ is the
energy in [$n-{1\over 2},n+{1\over 2}]$ for the even and odd
trial wavefunctions respectively, then the energy of a state
with m even pieces and n odd pieces is mF$_{e}+$nF$_{o}$. Thus
the total energy per interval is
(mF$_{e}+$nF$_{o})/($m$+$n). In particular, the energy per
interval is F$_{e}$ for the uniform chain of even pieces, i.e.
the ground state, and (mF$_{e}+$F$_{o}$)/(m$+1$) for the state
with only one odd piece, i.e. the state of Fig. 2. For an
$infinite$ number of spikes (m$\rightarrow\infty$), the two
states are $degenerate$,
as expected. Of course, the same holds for a state with
infinitely many even pieces, but only two odd pieces. If the
number of odd pieces becomes substantial though, then the energy
of the state will be definitely higher than that of the ground
state.

\vspace*{3mm}
For a large but finite number of spikes we still expect all these
various
states to be degenerate, as long as the minimum value of the even
pieces
is practically zero, because in that case $|\Psi|^{4}$ is
essentially the same for both even and odd pieces. We shall
verify this by explicit calculation, using our variational
wavefunctions. 

\vspace*{3mm}
We note that $\Psi_{en}(n)\rightarrow 0$ when
$\gamma\rightarrow\infty$. Therefore the degeneracy mentioned
above requires that $\gamma$ be very large. We shall neglect
therefore terms like sech$(\gamma/2)$. In this limit, 
\be
F_{e}\approx\nu\gamma|\psi|^{2}+{|\psi|^{4}\over 4\gamma}
+\int_{n-{1\over 2}}^{n+{1\over
2}}\,\,\,\,dz\,\,u(z)|\psi|^{2}
\frac{\cosh^{2}[\gamma(z-n)]}{\cosh^{2}(\gamma/2)}+O(e^{-\gamma})
\ee
\be
F_{o}\approx\nu\gamma|\psi|^{2}+{|\psi|^{4}\over 4\gamma}
+\int_{n-{1\over 2}}^{n+{1\over 2}}\,\,\,\,dz\,\,u(z)
|\psi|^{2}\frac{\sinh^{2}[\gamma(z-n)]}{\sinh^{2}(\gamma/2)}
+O(e^{-\gamma}).
\ee
Hence, since $\cosh^{2}(\gamma/2)\approx
\sinh^{2}(\gamma/2)\approx$$e^{\gamma}/4$,
\be
F_{o}-F_{e}\approx-\int_{n-{1\over 2}}^{n+{1\over 2}}\,\,\,\,dz\,
4e^{-\gamma}u(z)|\psi|^{2}.
\ee
And since $\gamma$ is large, and terms of order O($e^{-\gamma}$)
have been dropped in this calculation, Eq. (2.5) implies that
F$_{o}\approx$F$_{e}$. In other words,
if $\Psi_{en}(n)\approx 0$, then all the possible states are
practically degenerate, even for a finite large number of spikes,
because they consist of odd and even pieces
only, pieces which were shown to have the same energy. Note
that our results are very general so far. The only
restriction is that the $\gamma$ that minimizes Eq. (2.3) be
large. Our conclusions are valid though for $any$ u(z) that can
lead to a large $\gamma$.

\vspace*{3mm}
We illustrate the above general conclusions by restricting
ourselves now to the Kronig-Penney model:
\be
u(z)=1-\sum_{n}\alpha\delta(z-n-{1\over 2}).
\ee
This choice of u(z), where $\alpha$ is a positive constant,
implies that there is a periodic chain of deep quantum wells
along the z axis. 

\vspace*{3mm}
For this choice of u(z) then, and in the limit of large
$\gamma$, we get
\be
F_{o}\approx F_{e}\approx|\psi|^{2}\Bigl[\nu\gamma+{1\over
\gamma}-
\alpha\Bigr]+{|\psi|^{4}\over 4\gamma}.
\ee
Minimization with respect to $|\psi|^{2}$ gives
\be
|\psi|^{2}=2\gamma\Bigl[\alpha-\nu\gamma-{1\over \gamma}\Bigr]
\ee
and
\be
F_{o}\approx F_{e}\approx -\gamma\Bigl[\alpha-\nu\gamma-{1\over
\gamma}\Bigr]^{2}.
\ee
Minimization with respect to $\gamma$ yields
\be
\gamma=\frac{\alpha+\sqrt{\alpha^{2}+12\nu}}{6\nu},
\ee
or equivalently $3\nu\gamma-\alpha=1/\gamma$, in which case
$|\psi|^{2}=4\nu\gamma^{2}-4$. Since $\gamma\gg 1$, we shall have
\be
\gamma\approx\alpha/3\nu\gg 1.
\ee
Since $|\psi|^{2}\geq 0$, we must also have 
$\gamma\geq 1/\sqrt{\nu}$, which implies 
\be
\alpha^{2}\geq 4\nu.
\ee
$Whenever$ therefore $\nu$ and $\alpha$ satisfy the
$restrictions$ of Eqs. (2.11) and (2.12), we expect all the
possible states to
be essentially degenerate. In particular, the $highest$ excited
state, the one consisting of odd pieces only, is $degenerate$
with
the $ground$ state, which is a chain of even pieces. Note
furthermore that the wavefunction is nonzero only if
$\nu\leq\alpha^{2}/4$. When in fact $\nu=\alpha^{2}/4$, we have
a transition to a zero wavefunction, even though $\gamma$, which
takes then the value $1/\sqrt{\nu}$, may be quite large.

\vspace*{3mm}
As a numerical illustration, we choose the case $\nu=0.01$,
$\alpha=1$. Then the ground state and the highest excited state
(only even or only odd pieces respectively) have an energy
of $-$13.5207 in this variational model, with $\gamma=34.305$ and
$\psi=6.56$. The exact energy can be found using the methods
of section 3, and it is $-14.933$ for both the ground state and
the highest state, while $\psi=6.93$. So both calculations
indicate that all the states are degenerate, for this particular
choice of $\nu$ and $\alpha$.

\vspace*{3mm}
The example where u(z) is given by Eq. (2.6) will be
examined also in section 3, since it can be solved exactly. We
can generalize for $any$ simple oscillatory u(z) though, so long
as $\gamma$ is very large.

\section{Exact Solutions}

In this section we shall solve exactly the model of Eqs. (1.4)
and (2.6), verifying thus the variational results of the previous
section. We shall be interested in $those$ values of the
parameters
$\nu$ and $\alpha$ that yield excited states $almost$
$degenerate$ with the ground state.

\vspace*{3mm}
We should note that a large $\nu$ would imply that the kinetic
energy is dominant, making thus the wavefunction too stiff.
In other words, the ground state wavefunction would come as
closely as possible to a constant, a choice that minimizes the
kinetic energy. In that case the minimum value of the ground
state wavefunction would be far from zero.

\vspace*{3mm}
On the other hand, if $\nu$ were exactly zero, then the
wavefunction would follow the variations of u(z) exactly.
Hence we need a small value of $\nu$ if we are going to have an
excited state that is close in energy to the ground state, since
the wavefunction of such a state
varies dramatically between the spikes. Furthermore, if $\nu$
is zero the wavefunction will have arbitrary signs at the
wells, in which case the various excited states will all be
degenerate with the ground state. For small $\nu$, this
degeneracy will not be altered too drastically.

\vspace*{3mm}
We shall be interested therefore in the exact solutions of this
model, for small $\nu$. The energy functional is minimized
when 
\be
\nu\frac{\partial^{2}\Psi}{\partial z^{2}}=
\Bigl[\,1-\sum_{n}\,\alpha\,\delta(z-n-{1\over 2})\Bigr]\Psi
+|\Psi|^{2}\Psi
\ee
The solution $\Psi(z)$ will have periodic features similar to
those of u(z). Integrating Eq. (3.1) gives
the boundary condition for $\Psi(z)$
\be
-\alpha\Psi(n+{1\over 2})=
\nu\Bigl[\frac{\partial\Psi}{\partial z}(n+{1\over 2})_{+}
-\frac{\partial\Psi}{\partial z}(n+{1\over 2})_{-}\Bigr].
\ee
Thus $\Psi(z)$ has a kink at each spike of the potential, due to
the $\delta$-functions.

\vspace*{3mm}
Direct integration of Eq. (3.1) after multiplying it by
$\partial\Psi/\partial z$ gives the solution in each interval.
The ground state has no node, hence $\Psi$ will have a minimum
at the middle of each interval, while it will be symmetric around
each spike. Thus the $exact$ ground state is found to be
\be
\Psi(z)=\frac{q}{cn\Bigl[\sqrt{(1+q^{2})/\nu}\,\,(z-n),\,
(2+q^{2})/(2+2q^{2})\Bigr]},
\ee
for $n-{1\over 2}\leq z\leq n+{1\over 2}$, extended periodically
everywhere else. Here $cn$ is a Jacobi elliptic function, and
$q=\Psi(n)$ is the minimum value of $\Psi(z)$. The above
expression is valid for $any$ value of $\nu$, large or small, and
we can easily verify that it satisfies Eq. (3.1). 

\vspace*{3mm}
The boundary conditions of Eq. (3.2) require then that
\be
\alpha=2\nu\sqrt{\frac{1+q^{2}}{\nu}}\frac{sn\Bigl[
\sqrt{\frac{1+q^{2}}{4\nu}}, \frac{2+q^{2}}{2+2q^{2}}\Bigr]
dn\Bigl[\sqrt{\frac{1+q^{2}}{4\nu}},
\frac{2+q^{2}}{2+2q^{2}}\Bigr]}{cn\Bigl[
\sqrt{\frac{1+q^{2}}{4\nu}},\frac{2+q^{2}}{2+2q^{2}}\Bigr]},
\ee
where the $dn$ and $sn$ are also Jacobi elliptic functions. This
equation determines $q$ as a function of $\alpha$ and $\nu$. Note
that there is always a ground state, since we can always find an
appropriate $q$ for a given choice of $\nu$ and $\alpha$.

\vspace*{3mm}
In the limit of a wavefunction localized around the spikes of the
potential we expect $q$ to be small. Then Eq. (3.3)
reduces to
\be
\Psi(z)\approx q\cosh[(z-n)/\sqrt{\nu}].
\ee

This is precisely the solution of the linear Kronig
Penney model, as expected, since for very small q the nonlinear
terms become unimportant. If $q=0$, Eq. (3.4) yields
\be
\alpha=2\sqrt{\nu}\tanh(1/2\sqrt{\nu}).
\ee
So a nonzero ground state will exist only if $\alpha\geq
2\sqrt{\nu}
\tanh(1/2\sqrt{\nu})$. Note that for small $\nu$ this becomes the
restriction of Eq. (2.12), as it should. Furthermore, if $\alpha$
is close to its lower limit, then the change of sign of $\Psi$
for a first excited state will have to spread over a few more
spikes of the potential, and it
will not be restricted to just the region between two successive
spikes. In this
paper we shall not be concerned with this possibility, and we
shall restrict our attention to values of $\alpha$ far from the
lower bound of Eq. (3.6). Then the change of sign for the excited
states occurs within just one spacing. 

\vspace*{3mm}
Now the Jacobi elliptic function $cn(x,m)$ is a periodic
function, with roots at the odd multiples of the elliptic
function $K(m)$, where $K(m)=\int_{0}^{\pi/2}\,d\theta/
\sqrt{1-m\sin^{2}\theta}$. Indeed, $cn(0,m)=1$, $cn(K(m),m)=0$,
$cn(2K(m),m)=-1$, $cn(3K(m),m)=0$, $cn(4K(m),m)=1$.

\vspace*{3mm}
The ground state, as mentioned in the previous section, must
consist of even pieces everywhere, of the form given by Eq.
(3.3). It must therefore be positive everywhere, since it is
continuous. So the $\Psi$ of Eq. (3.3) must not be allowed to
become negative. This means that the quantity $\sqrt{(1+q^{2})
/\nu}|z-n|$ must be smaller than $K((2+q^{2})/(2+2q^{2}))$ within
the interval $[n-{1\over 2},n+{1\over 2}]$. Hence
\be
{1\over 2}\sqrt{(1+q^{2})/\nu}\leq K(\frac{2+q^{2}}{2+2q^{2}}).
\ee
This inequality holds for any values of $\alpha$ and $\nu$. If
$\nu$ is small, then the right hand side of this inequality has
to be large. This happens when the argument of $K(m)$ is close
to 1, in which case $K(m)\approx\ln\sqrt{16/(1-m)}$. In this
particular case, the argument is 1 if $q$ is very small. Thus if
$\nu$ is small, $q$ must be small. 

\vspace*{3mm}
Furthermore, when inequality (3.7) becomes an equality, the
denominator in Eq.
(3.3) tends to zero at $z=n+{1\over 2}$, and therefore the value
of $\Psi(z)$ at the spikes becomes infinite. In other words, the
value of $\Psi$ at the spikes can be much larger than the value
of $\Psi$ at the midpoints. 

\vspace*{3mm}
Indeed $\Psi(n-{1\over
2})/\Psi(n)=1/cn[\sqrt{(1+q^{2})/4\nu},(2+q^{2})/(2+2q^{2})].$
Hence, if $q\rightarrow 0$,
$\Psi(1/2)/\Psi(1)\approx\cosh(1/\sqrt{4\nu})$, which tends to
infinity when $\nu$ tends to zero. Thus the even pieces of $\Psi$
become very deep if $\nu$ is small, because in that case
$q\rightarrow 0$ and $\Psi(n\pm{1\over 2})$ is very large.

\vspace*{3mm}
We say in that case that the wells are weakly coupled. Note that
in that
case the $|\Psi|^{4}$ of an even piece would not differ too much
from the $|\Psi|^{4}$ of an odd piece. In other words, we expect
the various possible states to be very close in energy to the
ground state, as mentioned already in section 2. Indeed the case
$\nu=0$ would correspond to a complete decoupling of the values
of the wavefunctions at the spikes of the potential, and hence
to
a complete
degeneracy of all the various states.

\vspace*{3mm}
Let us examine more thoroughly the singularities that may arise
in the behavior of $\Psi$. We said that $\Psi(z)$ becomes very
large at the spikes of the potential when inequality (3.7)
becomes almost an
equality:
\be
\sqrt{(1+q^{2})/4\nu}\approx K(\frac{2+q^{2}}{2+2q^{2}}).
\ee
But if $m$ is very close to 1, then $K(m)\approx\ln\sqrt{16/(1-
m)}$. Hence this approximate equality reduces for small $q$ to
$1/2\sqrt{\nu}\approx\ln(\sqrt{32/q^{2}})$, whence
\be
q\approx\sqrt{32}\,e^{-1/2\sqrt{\nu}}.
\ee
Hence, if $\nu$ is small, and if the ground state wavefunction
has deep cups, we must have $q\approx
\sqrt{32}\exp(-1/2\sqrt{\nu})$. Note that
even though $q$ is small, the value of the wavefunction at the
minima of the potential is large.
Nonetheless, since $q\rightarrow 0$, the wavefunction is again
given by Eq. (3.5), an equation that tells us that $\Psi(z)$
falls to $1/e$ of its value within a distance of $\sqrt{\nu}$
from the spikes. In that sense we can say that the "thickness"
of $\Psi$ at each spike is $2\sqrt{\nu}$. But the peaks of $\Psi$
would overlap when the thickness of each peak equals the distance
between successive peaks. This happens when $\nu=1/4$. When we
speak therefore
of weakly coupled wells, we mean that $\nu\ll 1/4$. And it is
only such wells that can lead to an essentially degenerate
spectrum of states. 

\vspace*{3mm}
Let us then summarize our results for the ground state. There is
always a ground state, with $q=\Psi(n)$ being the minimum value
of $\Psi$ in the interval $[n-{1\over 2},n+{1\over 2}]$. This is
a symmetric series of even pieces, and it resembles a chain of
symmetric cups (see Fig. 3). The absolute value of the slope of
$\Psi$ at the layers is $\alpha\Psi(n+{1\over 2})/2\nu$. This
ground state will be degenerate with any other states if their
$|\Psi|^{4}$'s are approximately the same. This can happen only
if $q$ is almost zero, as explained in section 2, because then
the minimum of the $|\Psi|^{4}$ of the even piece approaches the
minimum of $|\Psi|^{4}$ of the odd piece, i.e. zero. But $q$ can
be tiny, and $\Psi$ still have a substantially nonzero value,
only close to the roots of the Jacobi elliptic finction $cn$ (see
Eq. (3.3)), i.e. for $q\approx\sqrt{32}e^{-1/2\sqrt{\nu}}$.
Furthermore, $q$ needs to be small in order to have the
degeneracy. Hence $\nu$ must be small. Indeed, the thickness
$2\sqrt{\nu}$ of each well implies that the wavefunctions around
the spikes will not overlap substantially, provided $\nu\ll 1/4$.
For small $\nu$ then we get a ground state which resembles a
chain of deep cups (see Fig. 3).

\vspace*{3mm}
Let us now proceed to the first excited state (see Fig. 2). Here
we assume again that $\nu$ is small, and hence the even pieces
will resemble deep cups. There will be only one odd piece, in the
interval $[-1/2,1/2]$, connecting a chain of negative even pieces
with a chain of positive even pieces. The characteristics of the
many even pieces will not be altered, because there
is only one odd piece. On the contrary, the characteristics of
the odd piece will be determined from those of the even pieces,
through the boundary conditions.

\vspace*{3mm}
Direct integration of Eq. (3.1) after multiplying it by
$\partial\Psi/\partial z$ will give the solution in the interval
[$-$1/2, 1/2], as long as we use the fact that $\Psi(0)=0$, since
there is one node there. For the first excited state there is
only one node, thus all the pieces outside the interval [$-$1/2,
1/2] will be even. 

\vspace*{3mm}
One can show thus that for $-1/2\leq z\leq 1/2$ we get the
$exact$ solution
\be
\Psi(z)=\sqrt{1-
\epsilon}\,\,\frac{sn[z\sqrt{(1+\epsilon)/2\nu},
2\epsilon/(1+\epsilon)]}{cn[z\sqrt{(1+\epsilon)/2\nu},
2\epsilon/(1+\epsilon)]},
\ee
with $\Psi^{\prime}(0)=\sqrt{(1-\epsilon^{2})/2\nu}$ and
$0\leq\epsilon\leq 1$. We can easily verify that this expression
satisfies Eq. (3.1). The value of $\Psi(z)$ at $z=1/2$, as
calculated from Eq. (3.10), must be equal to the one that can be
calculated from Eq. (3.3). This relation determines the parameter
$\epsilon$. If the even pieces are deep enough, i.e. if $\nu$ is
small enough, then the slope of $\Psi(z)$ at (1/2)$_{-}$ will
turn out to be $\alpha\Psi({1\over 2})/2\nu$.

\vspace*{3mm}
Indeed, we saw that if for small $\nu$ the value of $\Psi$ at the
spikes of the potential is very large, then
$q\approx\sqrt{32}e^{-1/2\sqrt{\nu}}$.
In general, the even and odd pieces correspond to the same energy
if $\Psi$ is very large at the spikes, $q$ being quite small. But
if $\Psi(1/2)$ is very large, then Eq. (3.10) implies that
$cn[\sqrt{(1+\epsilon)/8\nu},2\epsilon/(1+\epsilon)]\approx 0$,
so as to make $\Psi(z)$ almost diverge. Consequently
\be
\sqrt{(1+\epsilon)/8\nu}\approx K[2\epsilon/(1+\epsilon)],
\ee
where K is the elliptic function
$K(m)=\int_{0}^{\pi/2}\,d\theta/\sqrt{1-m\sin^{2}\theta}$. And
since $\nu$ is small, the value of $K$ will have to be rather
large, which means that $2\epsilon/(1+\epsilon)\rightarrow 1$,
i.e. $\epsilon\rightarrow 1$. Indeed, remembering that
$K(m)\approx\ln\sqrt{16/(1-m)}$ when $m\rightarrow 1$, we can
easily find that Eq. (3.11) is solved by the value
\be
1-\epsilon\approx 32e^{-1/\sqrt{\nu}}.
\ee
So this value of $\epsilon$ yields a very large $\Psi(1/2)$, for
small $\nu$. In fact, we must have in general, for any $\nu$,
\be
\sqrt{(1+\epsilon)/8\nu}\leq K[2\epsilon/(1+\epsilon)],
\ee
otherwise the elliptic function $cn$ would get a root in [0,1/2]
and $\Psi(z)$ would have a vertical asymptote there.

\vspace*{3mm}
We can now check the value of $\Psi(1/2)$. The second argument
of the elliptic functions $sn$, $dn$ and $cn$ is
$(2+q^{2})/(2+2q^{2})$ for the even pieces and
$2\epsilon/(1+\epsilon)$ for the odd pieces, both of which will
equal
$1-16e^{-1/\sqrt{\nu}}$ when Eqs. (3.9) and (3.12) hold, i.e. for
very large values of $\Psi(1/2)$. Therefore for very small $\nu$
this second argument is
essentially 1, in which case the $sn$ becomes $\tanh$, the $cn$
becomes $sech$, and the $dn$ $sech$. Then Eq. (3.10) gives
$\Psi(z)\approx\sqrt{1-\epsilon}\sinh(z\sqrt{(1+\epsilon)/2\nu})$
in $[-1/2,1/2]$, and $\Psi(z)\approx q\cosh[(z-n)/\sqrt{\nu}]$
in $[n-{1\over 2},n+{1\over 2}]$, where $n\ne 0$. Both
expressions give then the
same values for $\Psi(1/2)$ and $|\Psi^{\prime}(1/2)|$, as
expected.

\vspace*{3mm}
The procedure for finding the first excited state then consists
of finding the value of $\epsilon$ that would ensure continuity
of $\Psi(z)$ at $z=1/2$. In that case the slope at $z=(1/2)_{-}$
will turn out automatically to be the exact opposite of the slope
at $z=(1/2)_{+}$. Finding the ground state, on the other hand,
simply requires finding a $q$ such that $|\Psi^{\prime}(n+{1\over
2})|=$$(\alpha/2\nu)\Psi(n+{1\over 2})$. This relation is
precisely Eq. (3.4). 

\vspace*{3mm}
Finally, we can find the highest excited state, the one
consisting of odd pieces only, by extending periodically the odd
solution of Eq. (3.10), and finding a value of $\epsilon$ such
that $|\Psi^{\prime}(n+{1\over 2})|$$=(\alpha/2\nu)\Psi(n+{1\over
2})$. Since the minimum value $q$ of the even piece for the
solutions that interest us is
$\sqrt{32}e^{-1/2\sqrt{\nu}}$, i.e. practically zero, the energy
of the even piece and of the odd piece is essentially the same
since they have the same $-\int\,dz|\Psi|^{4}/2$. And all the
states are then degenerate.

\vspace*{3mm}
It is interesting to note that $\alpha=2\nu\Psi^{\prime}({1\over
2})/\Psi({1\over 2})$$\rightarrow 4\nu+{1\over 3}$ if
$\nu\rightarrow\infty$, for the $\Psi(z)$ of Eq. (3.10). Hence
the highest state does not exist when $\nu\rightarrow\infty$,
unless $\alpha=4\nu+{1\over 3}$. Similarly, quite a few other
excited states do not exist for large values of $\nu$. The even
ground state exists always though, if $\alpha$ is above the lower
bound of Eq. (3.6). We present numerical values
of the parameters and energies of the ground state, of the first
excited state and of the highest state in Tables 1, 2 and 3 for
various choices of $\nu$ and $\alpha$.

\vspace*{3mm}
Note that for a given choice of $\nu$, large or small, and a
given value of $\alpha$ greater than the lower bound of Eq.
(3.6), $q$ is given by Eq. (3.4). But it is only small $\nu$'s,
and $\alpha$'s quite far from their lower bound, that will lead
to a ground state with deep even pieces. In that case $q$ is
given by Eq. (3.9), and the ground state will be practically
degenerate with the first few excited states. 

\section{A Potential With Gaussian Wells}

We shall examine numerically the bound states that correspond to
the
choice
\be
u(z)=\,\,1\,-\,g\sum_{n}\exp[-b(z-n-{1\over 2})^{2}],
\ee
shown in Fig. 4. We choose the parameters so that $e^{b/4}>g>1$.
Then u(z) is negative at its minima and positive at the
midpoints between the minima.
Therefore, if $\nu$ were exactly zero, then there would be
a nonzero wavefunction at the minima of u(z), but there would be
a region
around the midpoints where $\Psi$ would be exactly zero. When
$\nu$ is slightly positive, the stiffness of the wavefunction
makes $\Psi$ leak into the "forbidden" regions (tunnelling). If
however $b<4\ln g$, then there is no tunnelling. We need a large
enough $b$ in order to get tunnelling.

\vspace*{3mm}
If on the other hand $b$ is extremely large, while $g$ remains
finite, the width of each well is reduced,
its strength remaining unaltered. We expect therefore that for
a given $g$ we cannot increase $b$ indefinitely, because we shall
not be able to find a solution. The numerical calculations do
indeed verify this.

\vspace*{3mm}
Thus, if $b$ is large enough, but not too large, and if $g$ is
large enough to allow a nonzero solution, we shall have
a competition between the kinetic energy and the rest of the
energy. The
latter forces $\Psi$ to follow the variations of
u(z). The kinetic energy, on the other hand, wants
$\Psi$ to be constant in space. Solutions such as the one of Fig.
2 will
be relevant for small $\nu$, since large values of
$\nu$ tend to push $\Psi$ towards a constant. Indeed, numerical
calculations verify that for large $\nu$ (e.g. $\nu=0.3$), the
kinetic energy is strong enough to force $\Psi_{max}$ to be close
to $\Psi_{min}$, and $\Psi_{min}\gg 0$. The ground state is then
a chain of shallow little cups away from zero.
 
\vspace*{3mm}
Excited states that will be almost degenerate in energy with
the ground state will appear at small values of $\nu$, when the
minimum of the ground state wavefunction between the wells
is very small, while $\Psi_{max}\gg 0$. Fig. 5 shows the ground
state, the first excited state, and the highest excited state,
for $\nu=0.01,$ with the corresponding energies per interval, for
a sample of 20 gaussian wells. Fig. 6 shows the
same states, but for $\nu=0.05$. We see again the characteristics
mentioned earlier: full degeneracy if $\nu$ is small enough, in
which case the minimum of $\Psi$ between wells is quite small
compared to its value $\Psi_{max}$ at the wells. The value of
$\Psi_{max}$ depends strongly on $g$. 

\vspace*{3mm}
Remember also that the energy difference between the ground state
and the first excited state is strictly equal to zero for an
infinite number of wells. This is due to the fact that the
wavefunction of the first excited state is exactly equal or
exactly
opposite to the wavefunction of the ground state on almost all
of the infinitely many wells. Indeed, in Figs. 7 and 8 we see
the first excited state for a fairly large value of $\nu$, and
a series of values for $b$ and $g$. Here the stiffness of the
wavefunction is quite large, so the change of the sign cannot
take place within just one interval. This change now occurs over
three or four wells. For small values of $\nu$ though, when the
wavefunction is quite malleable, the wavefunction changes sign
within one interval. 

\section{Conclusions}

We have studied the bound states of a nonlinear version of the
Schrodinger equation for the Kronig Penney model, a version that
is relevant to quite a few phenomena in condensed matter physics.

\vspace*{3mm}
We have seen that there is a substantial range of
parameters, not just for the Kronig Penney model but for other
simple oscillatory choices of u(z) as well, for which
the various states are essentially degenerate in
energy. This degeneracy requires that $\nu$ be small enough, so
as to allow the wavefunction to have a small, but 
non-negligible, value at the midpoints between the wells. At the
same time the value of the wavefunction at the wells can be
quite substantial. Then there is very little cost in having
$\Psi$ change
sign in going from one well to the next. In fact, the energy
differences are really small when $\Psi_{max}\gg\Psi_{min}$. This
means that the lower excited states, which connect regions of
positive
$\Psi$ with regions of negative $\Psi$, become as favorable as
the ground state. This degeneracy is exact in the limit of
infinitely many quantum wells for these lower excited states.
Even the highest excited state though, which consists of odd
pieces only, has an energy very close to the energy of the ground
state for sufficiently small $\nu$. 

\vspace*{3mm}
We demonstrated this basic idea in Section 2 through a
variational calculation valid for a generic oscillatory
potential. The variational results were confirmed through the
exact solution of the nonlinear Kronig Penney model, presented
in Section 3, as well as through a numerical calculation for the
case of another simple oscillatory potential, presented in
Section 4. All these calculations show the exact (for infinitely
many spikes) or approximate (for finitely many spikes) degeneracy
of the ground state.

\newpage

\newpage 
\noindent
{\bf Table Captions \hfill}
 
\begin{enumerate}

\item[\bf Table 1:] 
Parameters of the ground state for $\delta$-function wells, for
various choices of $\nu$ and
$\alpha$. $E_{g}$ is the energy per interval. \\
\begin{center}
\begin{tabular}{|c|c|c|c|c|c|}
\hline
$\nu$ & $\alpha$ & $\Psi(1)=q$ & $\Psi(1/2)$ & $-
\Psi^{\prime}(1/2)_{+}$ & $E_{g}$ \\
\hline
0.002 & 1    & 7.2x10$^{-5}$ & 15.748  & 3937    & $-$81.4526 \\
0.01  & 0.25 & 0.0127        & 1.0608  & 13.2601 & $-$0.02710 \\
0.01  & 1    & 0.031074      & 6.92821 & 346.41  & $-$14.9334 \\
0.01  & 10   & 0.037283      & 70.6965 & 35348.3 & $-$16646.9 \\
0.05  & 1    & 0.34647       & 2.83402 & 28.3402 & $-$1.95903 \\
0.05  & 10   & 0.511176      & 31.5911 & 3159.11 & $-$3314    \\
0.1   & 1    & 0.497576      & 1.78192 & 8.90962 & $-$0.59964 \\
\hline
\end{tabular}
\end{center}

\newpage

\item[\bf Table 2:]
Parameters of the first excited state for $\delta$-function
wells, for various choices of
$\nu$ and $\alpha$. $E_{1}$ is the energy in [$-1/2,1/2]$.
For the other intervals, the energy is the $E_{g}$ given in
Table 1.\\
\begin{center}
\begin{tabular}{|c|c|c|c|c|c|c|}
\hline
$\nu$ & $\alpha$ & $q$ & $\epsilon$ & $\Psi(1/2)$ &
$\Psi^{\prime}(1/2)_{-}$ & $E_{1}$ \\
\hline
0.002 & 1    & 7.2x10$^{-5}$ & 1-5.2x10$^{-9}$ & 15.748 & 3937 
& $-$81.4526  \\
0.01  & 0.25 & 0.0127        & 0.999838        & 1.0608 & 13.2613
& $-$0.027094 \\
0.01  & 1    & 0.031074      & 0.99903         & 6.9282 & 346.41
& $-$14.9333  \\
0.01  & 10   & 0.037283      & 0.998598        & 70.696 & 35348.3
& $-$16646.9  \\
0.05  & 1    & 0.34647       & 0.82356         & 2.8340 & 28.4417
& $-$1.92445  \\
0.05  & 10   & 0.511176      & 0.363395        & 31.591 & 3159.12
& $-$3313.85  \\
0.1   & 1 & 0.497576         & 0.424940        & 1.7819 & 9.28766
& $-$0.526821 \\
\hline
\end{tabular}
\end{center}

\newpage 
\item[\bf Table 3:]
Parameters of the highest state for $\delta$-function wells, for
various choices of $\nu$ and
$\alpha$. $E_{h}$ is the energy per interval.\\
\begin{center}
\begin{tabular}{|c|c|c|c|c|c|}
\hline
$\nu$ & $\alpha$ & $\epsilon$ & $\Psi(1/2)$ &
$\Psi^{\prime}(1/2)_{-}$ & $E_{h}$ \\
\hline
0.002 & 1    & 1-5.2x10$^{-9}$ & 15.748 & 3937    & $-$81.4526
\\
0.01  & 0.25 & 0.999838        & 1.0605 & 13.2565 & $-$0.02707
\\
0.01  & 1    & 0.99903         & 6.9282 & 346.41  & $-$14.9332
\\
0.01  & 10   & 0.998598        & 70.696 & 35348.3 & $-$16646.9
\\
0.05  & 1    & 0.82442         & 2.8213 & 28.213  & $-$1.89582
\\
0.05  & 10   & 0.363395        & 31.591 & 3159.11 & $-$3313.84
\\
0.1   & 1    & 0.516054        & 1.6527 & 8.26345 & $-$0.40157
\\
1     & 4.34 & 0.730536        & 0.2518 & 0.54642 & $-$0.00038
\\
\hline
\end{tabular}
\end{center}

\end{enumerate}

\newpage
\noindent
{\bf Figure Captions \hfill}
 
\begin{enumerate}

\item[\bf Figure 1:] 
(a) A typical wavefunction when the potential wells are very far
apart
($\nu\approx 0$).

(b) A typical wavefunction when the potential wells are closer
together
($\nu\ll 1)$.

\item[\bf Figure 2:] 
First excited state with the root at $z=0$, for $\delta$-function
wells.

\item[\bf Figure 3:] 
The ground state, for $\delta$-function wells.

\item[\bf Figure 4:]
The potential u(z) for gaussian wells ($g=2.5$, $b=20$).

\item[\bf Figure 5:]
The ground state, the first excited state, and the highest state,
for 20 gaussian wells with $\nu=0.01$, g=2.5, b=20, with the
corresponding
energies per interval.

\item[\bf Figure 6:]
The ground state, the first excited state, and the highest state,
for 20 gaussian wells with $\nu=0.05$, g=2.5, b=20, with the
corresponding
energies per interval.

\item[\bf Figure 7:]
First excited state for gaussian wells with $\nu=0.1$, $b=20$,
and
$g$=2.2, 2.5 and 5.

\item[\bf Figure 8:]
First excited state for gaussian wells with $\nu=0.1$, $g=2.5$
and
$b=$5, 10, 20, and 25.

\end{enumerate}

\begin{thebibliography}{99}
\bibitem{g1}

T. Koyama, N. Takezawa, Y. Naruse, M. Tachiki, Physica
{\bf C 194}, 20 (1992); T. Koyama, M. Tachiki, Physica {\bf C
193}, 163 (1992); M. Machida, H. Kaburaki, Physical Review
Letters {\bf 74}, 1434 (1995); L. Balents, L. Radzihovsky,
Physical Review Letters {\bf 76}, 3416 (1996).

\bibitem{g2}
S. Ami, K. Maki, Progress of Theoretical Physics {\bf
53}, 1 (1975).

\bibitem{g3}
J. Betouras, R. Joynt, Z. W. Dong, T. Venkatesan, P. Hadley,
Applied Physics Letters {\bf 69}, 2432 (1996).

\bibitem{g4}
J. Betouras, R. Joynt, Physica {\bf C 250}, 256 (1995). 

\bibitem{g5}
I. Khlyustikov, A. Buzdin, Advances in Physics {\bf
36}, 271 (1987); A. Abrikosov, A. Buzdin, M. Kulic, D. Kuptsov,
Soviet Physics JETP {\bf 68}, 210 (1989).

\bibitem{g6}
K. Hizanidis, D. Frantzeskakis, C. Polymilis, Journal
of Physics {\bf A29}, 7687 (1996); D. Frantzeskakis, Journal of
Physics {\bf A29}, 3631 (1996); H. Frauenkron, P. Grassberger,
Journal of Physics {\bf A28}, 4987, (1995).  



\end{thebibliography}
\end{document}